\documentclass[reprint, amsmath, amssymb, superscriptaddress, footinbib]{revtex4-2}

\usepackage{graphicx}
\usepackage{amsmath}
\usepackage{natbib}
\usepackage{caption}
\usepackage{float}
\usepackage{dcolumn}
\usepackage{bm}
\usepackage[colorlinks=true,citecolor=blue,linkcolor=blue,urlcolor=blue]
{hyperref}
\usepackage{subfig}

\begin{document}
	
	\title{Wormhole formation in massive gravity: An analytic description}
	
	
	\author{Ayanendu Dutta}
	\email{ayanendudutta@gmail.com}
	\affiliation{Department of Physics, Jadavpur University, Kolkata-700032, India}
	
	\author{Dhritimalya Roy}
	\email{rdhritimalya@gmail.com}
	\affiliation{Department of Physics, Jadavpur University, Kolkata-700032, India}
	
	\author{Nihal Jalal Pullisseri}
	\email{nihal.jalaluddin@gmail.com}
	\affiliation{Department of Physics, St. Stephen's College, Delhi-110007, India}

	\author{Subenoy Chakraborty}
	\email{schakraborty.math@gmail.com}
	\affiliation{Department of Mathematics, Jadavpur University, Kolkata-700032, India}

	
\begin{abstract}
	The present study analyses the wormhole solution both in the dRGT-$ f(R,T) $ massive gravity and Einstein massive gravity. In both the models, the anisotropic pressure solution in ultrastatic wormhole geometry gives rise to the shape function that involves massive gravity parameters $ \gamma $ and $ \Lambda $. However, the terms consisting of $ \gamma $ and $ \Lambda $ acts in such a way that the spacetime loses asymptotic flatness. Similar to the black hole solution in massive gravity, this inconsistency arises due to the repulsive effect of gravity which can be represented by the photon deflection angle that goes negative after a certain radial distance. It is investigated that the repulsive effect induced in the massive gravitons push the spacetime geometry so strongly that the asymptotic flatness is effected. On the other hand, in this model, one can have a wormhole with ordinary matter at the throat that satisfies all the energy conditions while the negative energy density is sourced by massive gravitons. Finally, using the TOV equation, it is found that the model is stable under the hydrostatic equilibrium condition.
\end{abstract}

\keywords{Wormhole solution, dRGT massive gravity, Non-asymptotic flatness, Repulsive gravity effect}

\maketitle

\section{Introduction}
Wormholes are the smooth bridges between two different universes, or sometimes between two distant parts of the same universe. The concept was first put forward by Einstein and Rosen in their famous Einstein-Rosen bridge in 1935 \cite{einstein-rosen}, where Misner and Wheeler first coined the term `Wormhole' later in 1957 \cite{misner1957}. The exact solution of the Einstein Field Equation for traversable wormhole were first successfully examined much later by Morris and Thorne in the year 1988 \cite{morris1988,morris1988_2}. They investigated that the energy-momentum component for such a wormhole always violates the null energy condition \cite{morris1988, visser_lorentzian-wormhole} which is the weakest classical energy condition that in turn violates all other energy conditions. So, for the construction of traversable wormhole one needs a matter with negative energy density called the exotic matter. For example, one may consider the well-known Ellis wormhole solution of General Relativity for instance \cite{ellis1973,ellis1979,bronnikov1973,kashargin2008,kashargin2008_2,kleihaus2014,chew2016}. However, if traversable wormholes are constructed in modified theories of gravity, the requirement of exotic matter is highly reduced. There are numerous studies that deals with the wormhole geometries and their corresponding energy conditions in various modified gravities. For example, one may go through the wormhole solutions in $ f(R) $ gravity \cite{lobo2009,mishra2022}, in $ f(R,T) $ gravity \cite{moraes2017,yousaf2017,elizalde2018,sharif2019}, and in other theories \cite{sc1,sc2,sc3,sc4,sc5,sc6,sc7,sc8,sc9,sc10,sc11,sc12,sc13,sc14,sc15}.

Ever since the traversable wormhole model was presented, it was a point of interest if wormholes can be constructed with ordinary matter. Recently, a study has been introduced to discuss the wormhole solutions both in Einstein gravity and in modified gravity theories, and it is established that we may have a wormhole with ordinary matter in modified gravity, which may satisfy all the energy conditions \cite{epl_paper}. Although, the usual matter can be ordinary, but the effective geometric matter which is the matter source of modified gravity violates the usual null energy condition. There are various studies conducted on the ground that obtained non-exotic matter wormholes \cite{fukutaka1989,hochberg1990,ghoroku1992,furey2005,bronnikov2010,kanti2011,kanti2012,harko2013,moraes2018,godani2021,sengupta2022}.

There are models of modified gravity theory as an extension of Einstein's GR to address cosmological phenomenon. The massive gravity theory is an important candidate in this context. The discovery of gravitational waves due to merger of two Black holes and massive stars as detected by LIGO and VIRGO \cite{abbott2016,abbott2017}, there is a severe constraint in the mass of the graviton, and thus, massive gravity theory has gained significant interest in theoretical regime. The de Rham-Gabadadze-Tolley (dRGT) massive gravity \cite{derham2010,derham2011} is an interesting nonlinear generalization to overcome the so-called van Dam-Veltman-Zakharov discontinuity \cite{dam1970,zakharov1970} of Fierz-Pauli linear massive gravity \cite{fierz1939}. The dRGT theory is utterly promising as it avoids the long standing issue of Boulware-Deser ghost instability \cite{boulware1972} that usually arises on the introduction of nonlinear generalization of massive gravity theory. In the simplest dRGT model of static spherically symmetric solution, there arises two additional characteristics scales (i.e., $ \gamma $ and $ \Lambda $) as compared to the Schwarzschild solution of General Relativity \cite{ghosh2016}. In the theory, the massive graviton generates these parameters (one of which is the effective cosmological constant) \cite{gumrukcuoglu2011,gumrukcuoglu2012} which can be used to address the dark matter and dark energy problems of the galactic and extragalactic scenario in a single theory. The nonlinear graviton interaction generating density and pressure which can behave as dark energy, and as a result, the galaxy rotation curve is influenced due to generation of dark matter halo. Thus, the massive gravitons may act as the dark matter halo resulting in asymptotically flat rotation curves. In \cite{panpanich2018}, the dRGT model is quite successfully fitted with observational data for Milky Way rotation curves, and LSB galaxies without considering any additional dark matter. The consistency of dRGT model with the NFW profile has also been noted in this context.
	
Further, the massive gravitons play the role of an anisotropic fluid, which in turn is a kind of dark energy. As a consequence, one may able to explain the late time cosmic acceleration, although, it can not justify the past inflation. Subsequently, in a rectified framework, it is found that the massive gravity theory is consistent \cite{afshar2022} with the Plank 2018 data \cite{plank2018} and its combination with BK18 and BAO \cite{bk18}.

The phenomenology in dRGT massive gravity for compact objects is interestingly promising. Various studies have already been conducted on Black hole, Black string, rotating Black string solutions, stability and greybody factor on charged Black hole and Black strings in dRGT model which are modified significantly from the conventional studies due to the presence of massive gravitons \cite{ghosh2016_2,tannukij2017,ghosh2020,nieuwenhuizen2011,burikham2017,ponglertsakul2018,boonserm2018,boonserm2019}.

On the other hand, the existence of wormhole demands the violation of null energy condition. At the same time, a natural outcome in the massive graviton density is that it can be negative in a certain region where the radial pressure is positive. So, in the massive gravity theory, the massive graviton energy-momentum tensor naturally fulfil the feature of energy condition violation \cite{sushkov2015}. Thus, the existence of wormhole in the massive gravity can be a natural outcome, and the discussion of this particular object has an additional importance on its own. There are only couple of studies present on wormhole solutions in dRGT massive gravity with its own limitations. Tangphati {\it et. al.} investigated the traversable wormhole solution in $ f(R) $ massive gravity \cite{tangphati2020}, where they considered an exponential shape function in order to discuss the asymptotic geometry. But, due to their restricted choices of shape function, they were unable to obtain a geometry which includes the effect of massive gravitons. Alongside, Kamma {\it et. al.} presented the wormhole solution in conventional Einstein gravity on the background of massive gravitons \cite{kamma2021}, and obtained both the redshift function and shape function which includes massive gravity parameters. Although, they discussed about negative pressure tension in wormholes, but they didn't shed any light on the asymptotic geometry triggered by the redshift and shape function.

So, in this article, we shall investigate the wormhole solution in $ f(R,T) $ massive gravity (the field equations are discussed in section \ref{field_eq}), and discuss the corresponding anisotropic pressure solution in the ultrastatic wormhole geometry in section \ref{aniso}. It is observed that a repulsive gravity effect is present and the asymptotic structure is violated in the model which is not so well-known in wormhole geometries. Hence, to represent the relationship between asymptotic structure and the repulsive gravity in a more generalized scenario, we examined wormhole solution in dRGT extension of Einstein gravity in section \ref{einstein}. On the other hand, the stability of the model is analyzed in section \ref{stability} followed by the discussions of energy conditions and matter content in section \ref{en_con}. Finally, section \ref{discuss} is dedicated for the conclusions and discussions.

We consider the natural units throughout the study, i.e., $ G=c=1 $.

\section{The Field equations}\label{field_eq}
To investigate the $ f(R,T) $ gravity model in the context of de Rham-Gabadadze-Tolley (dRGT) massive gravity, we consider the action in the following form
\begin{equation}\label{eq1}
	S = \int d^4 x \sqrt{-g} \bigg(\frac{1}{16\pi}\Big[ f(R,T) + m^2_g \mathcal{U}(g,\phi^a) \Big] +\mathcal{L}_{m} \bigg),
\end{equation}
where $ f(R,T) $ is function of $ R $ and $ T $, $ \mathcal{U} $ is the self-interacting potential of the graviton with graviton mass $ m_g $, $ \mathcal{L}_{m} $ is the matter Lagrangian, and $ g $ is the determinant of the metric tensor $ g_{\mu \nu} $. Here, the potential $ \mathcal{U} $ is defined as
\begin{eqnarray}\label{eq2}
	\mathcal{U} = \mathcal{U}_2 + \alpha_3 \mathcal{U}_3 + \alpha_4 \mathcal{U}_4 .
\end{eqnarray} 
where $\mathcal{U}_{2},\,\mathcal{U}_{3}$ and $\mathcal{U}_{4}$ are given by
\begin{eqnarray}\label{eq3}
	\mathcal{U}_2 &=& [\mathcal{K}]^2 - [\mathcal{K}^2], \nonumber \\
	\mathcal{U}_3 &=& [\mathcal{K}]^3 - 3[\mathcal{K}][\mathcal{K}^2] + 2 [\mathcal{K}^3], \nonumber \\
	\mathcal{U}_4 &=& [\mathcal{K}]^4 - 6[\mathcal{K}]^2[\mathcal{K}^2] + 8[\mathcal{K}][\mathcal{K}^3] + 3[\mathcal{K}^2]^2 -     6[\mathcal{K}^4], \nonumber \\
	\mathcal{{K}^{\mu}}_{\nu} &=& \delta^{\mu}_{\nu} - \sqrt{g^{\mu\lambda} \partial_{\lambda}\phi^a \partial_{\nu}\phi^b \mathcal{F}_{ab} },
\end{eqnarray}

Here, $ [\mathcal{K}] $ represents the trace of $ \mathcal{K}^\mu_\nu $, where $ (\mathcal{K}^i)^{\mu}_{\nu} = \mathcal{K}^\mu_{\rho_1} \mathcal{K}^{\rho_1}_{\rho_2} ...\mathcal{K}^{\rho_i}_\nu $, $ \phi^a $ is the St\"uckelberg field, and the reference fiducial metric $ \mathcal{F}_{ab} $ has the explicit form given by
\begin{equation}\label{eq4}
	\mathcal{F}_{ab} = 
	\begin{pmatrix}
		0 & 0 & 0 & 0 \\
		0 & 0 & 0 & 0 \\
		0 & 0 & c^2 & 0 \\
		0 & 0 & 0 & c^2 \text{sin}^2 \theta \\
	\end{pmatrix},
\end{equation}
where the unitary gauge is fixed as, $ \phi^a = x^\mu \delta^a_\mu $, so that
\begin{equation}\label{eq5}
	\sqrt{g^{\mu\lambda} \partial_{\lambda}\phi^a \partial_{\nu}\phi^b \mathcal{F}_{ab} }= \sqrt{g^{\mu \lambda} \mathcal{F}_{\lambda \nu} }.
\end{equation}

Now, by varying the action with respect to the metric $ g_{\mu \nu} $, we arrive to the field equation for modified dRGT-$ f(R,T) $ massive gravity theory given by
\begin{eqnarray}\label{eq6}
\nonumber	&&f_R(R,T) R_{\mu \nu} -\frac{1}{2} f(R,T) g_{\mu \nu}\\
&& +(g_{\mu \nu} \Box -\nabla_\mu \nabla_\nu) f_R(R,T)\\ &&= -m^2_g X_{\mu \nu} + 8\pi T_{\mu \nu} -f_T(R,T)(T_{\mu \nu}+\Theta_{\mu \nu}), \nonumber
\end{eqnarray}
where $ f_R(R,T) $ and $ f_T(R,T) $ are the differentiation of $ f(R,T) $ with respect to $ R $ and $ T $ respectively, and $ \Box f_R= g^{\mu \nu} \nabla_\mu \nabla_\nu f_R $. We also have the variation of trace of energy-momentum tensor of the matter field, $ T= g^{\mu \nu} T_{\mu \nu} $ as
\begin{equation}\label{eq7}
	\frac{\delta(g^{\alpha \beta} T_{\alpha \beta})}{\delta g^{\mu \nu}} =T_{\mu \nu} +\Theta_{\mu \nu},
\end{equation}
where $ \Theta_{\mu \nu} $ and $ T_{\mu \nu} $ are given by
\begin{eqnarray}\label{eq8}
	\Theta_{\mu \nu} \equiv g^{\alpha \beta} \frac{\delta T_{\alpha \beta}}{\delta g^{\mu \nu}},
	\label{eq8}\\
	T_{\mu \nu} \equiv g_{\mu \nu} \mathcal{L}_m - \frac{2 \partial(\mathcal{L}_m)}{\partial g^{\mu \nu}}.
	\label{eq9}
\end{eqnarray}

Here, $ T_{\mu \nu} $ can also be written in the principal pressure terms as
\begin{equation}\label{eq10}
	T_{\mu \nu}= (\rho+p_t)u_\mu u_\nu +p_t g_{\mu \nu} +(p_r-p_t)\chi_\mu \chi_\nu,
\end{equation}
where $ u_\mu $ is the timelike unit vector, $ \chi_\mu $ is a spacelike unit vector orthogonal to timelike unit vector, such that $ u_\mu u^\mu=-1 $ and $ \chi_\mu \chi^\mu=1 $. Assuming the universal choice of Lagrangian matter density $ \mathcal{L}_m=\rho $, we have $ \Theta_{\mu \nu}= -2T_{\mu \nu}+\rho g_{\mu \nu} $.

Alongside, $ \chi_{\mu \nu} $ is the massive graviton tensor, given by
\begin{widetext}
	\begin{eqnarray}\label{eq11}
		 \nonumber	X^\mu_\nu &=& \mathcal{K}^\mu_\nu -[\mathcal{K}] \delta^\mu_\nu -\alpha \left[ (\mathcal{K}^2)^\mu_\nu -[\mathcal{K}]\mathcal{K}^\mu_\nu +\frac12 \delta^\mu_\nu \left( [\mathcal{K}]^2 -[\mathcal{K}^2] \right) \right] \\
		&&+3\beta \left[ (\mathcal{K}^3)^\mu_\nu -[\mathcal{K}](\mathcal{K}^2)^\mu_\nu +\frac12 \mathcal{K}^\mu_\nu \left( [\mathcal{K}]^2- [\mathcal{K}^2] \right) \right] -3\beta \left[ \frac16 \delta^\mu_\nu \left( [\mathcal{K}]^3- 3[\mathcal{K}] [\mathcal{K}^2]+ 2 [\mathcal{K}^3] \right) \right],
	\end{eqnarray}
\end{widetext}
where the parameters $ \alpha $ and $ \beta $ are defined by
\begin{eqnarray}\label{eq12}
	\alpha = 1 + 3\alpha_3\,,\qquad \beta = \alpha_3 + 4\alpha_4.
\end{eqnarray}

According to the definition,
\begin{equation}\label{eq13}
	\frac{m^2_g}{8\pi}X_{\mu \nu}= -(\rho^{(g)}+p_t^{(g)})u_\mu u_\nu -p_t^{(g)} g_{\mu \nu} -(p_r^{(g)}-p_t^{(g)})\chi_\mu \chi_\nu,
\end{equation}
which is nothing but the energy momentum tensor of the massive gravity sector. Using Eq. \eqref{eq11}, one can readily calculate the density and pressure components $ \rho^{(g)}(r) $ and $ p^{(g)}_{r,\perp}(r) $ which are given by \cite{burikham2016,kareeso2018,panpanich2018,tangphati2020}
\begin{eqnarray}
	\rho^{(g)}(r) &\equiv& \frac{m_g^2}{8\pi}{X^t}_t 
	= -\frac{1}{8\pi}\left( \frac{2 \gamma -  \Lambda r}{r} \right),
	\label{eq14}\\
	p_r^{(g)}(r) &\equiv& -\frac{m_g^2}{8\pi}{X^r}_r = \frac{1}{8\pi}\left( \frac{2 \gamma -  \Lambda r}{r} \right),
	\label{eq15}\\
	p_{\theta, \phi}^{(g)}(r) &\equiv& -\frac{m_g^2}{8\pi}X_{\theta,\phi}^{\theta,\phi} 
	= \frac{1}{8\pi}\left( \frac{ \gamma -  \Lambda r}{r} \right),
	\label{eq16}
\end{eqnarray}
where effective cosmological constant $ \Lambda $ and a new parameter $ \gamma $ are introduced and written in the linear combination of the parameters $ \alpha $ and $ \beta $, given by
\begin{eqnarray}\label{eq17}
	\Lambda \equiv -3m_g^2(1+ \alpha + \beta), \quad \gamma \equiv -m_g^2c(1 + 2\alpha + 3\beta).
\end{eqnarray}

Here at this point, after applying some simple mathematical calculations on Eq. \eqref{eq6}, we arrive to the final field equation.
\begin{equation}\label{eq18}
	G_{\mu \nu}=  8\pi G_{eff} T_{\mu \nu}+ T^{eff}_{\mu \nu} -\frac{1}{f_R(R,T)}m_g^2 X_{\mu \nu},
\end{equation}
where
\begin{eqnarray}
	G_{eff} &=& \frac{1}{f_R(R,T)} \left( 1+\frac{f_T(R,T)}{8\pi} \right), \label{eq19} \\
	T^{eff}_{\mu \nu} &=& \frac{1}{f_R(R,T)} \Big[ \frac12 (f(R,T)-R f_R(R,T) \\ \nonumber
	 &+& 2\rho f_T(R,T)) g_{\mu \nu} -(g_{\mu \nu} \Box -\nabla_\mu \nabla_\nu) f_R(R,T) \Big].
	 \label{eq20}
\end{eqnarray}

Note that, when $ f(R,T) \equiv f(R) $, such that $ f_T(R,T)=0 $, we get back the usual $ f(R) $ massive gravity solution \cite{tangphati2020}. In this field equation, the energy-momentum tensor component ($ T_{\mu \nu} $) of $ f(R,T) $ gravity represents the interaction between matter and curvature, and one may interpret this as the curvature-matter coupling occurs due to the exchange of energy and momentum between the both. Otherwise, the total energy-momentum tensor which is the sum of $ f(R,T) $ and massive gravity sectors, adds the interaction of massive gravitons with the curvature-matter coupling. It can be expressed as
\begin{equation}\label{eq22}
	T^{tot}_{\mu \nu}= diag\left( -\rho-\rho^{(g)}, p_r+p_r^{(g)}, p_t+p_t^{(g)}, p_t+p_t^{(g)} \right).
\end{equation}

However, $ f(R,T) $ gravity proposed by Harko \textit{et al.} \cite{harko2011} has a critical issue that violates the energy conservation law, and it leads to non-geodesic motion of test particles. But, an alternative approach of $ f(R,T) $ gravity proposed by Chakraborty \cite{chakraborty2013} showed that the form of the field equation remains similar if one takes into account of the conservation of EM tensor. As a result, test particles move in geodesic orbits, and the choice of Lagrangian is not completely arbitrary. For homogeneous and isotropic model of the universe, the approach refers to the field equation which is equivalent to the Einstein gravity with non-interacting 2-fluid system, one of which is the usual perfect fluid in modified gravity and the other one shows exotic nature.

Recently, in \cite{ilyas2022}, it is showed that in $ f(R,T) $ gravity, one may have a wormhole with non-exotic matter where the ordinary EM tensor satisfies the NEC, and the extra curvature caused by the modified gravity manipulate violation of NEC, so that the total EM tensor violates NEC. In a general perspective of modified gravity, the existence of ordinary matter wormhole is briefly discussed in \cite{epl_paper}. For another example of non-exotic matter wormhole in $ f(R,T) $ gravity, readers are referred to \cite{banerjee2021}.

Now, it is mathematically evident that the massive gravity EM tensor holds the property of anisotropic dark energy, as $ p_r^{(g)} = -\rho^{(g)} $, which may successfully manipulate the phenomenological aspects of various compact objects and astrophysical phenomenons \cite{ghosh2016_2,tannukij2017,ghosh2020,nieuwenhuizen2011,burikham2017,ponglertsakul2018,boonserm2018,boonserm2019,panpanich2018}. Thus, in $ f(R,T) $-massive gravity, the massive gravity and curvature-matter coupled system has the massive gravitons and effective geometric matter (due to the curvature) that can exhibit exotic nature. Hence, if the EM sector of usual matter satisfies the energy conditions and is dominated by the sum of gravitons and geometric matter, we can have a usual matter wormhole.

Now, to construct traversable wormhole solution, we consider the Morris-Thorne line element given by \cite{morris1988}
\begin{equation}\label{eq23}
	ds^2= -e^{2 \Phi(r)}dt^2+\left( 1- \frac{b(r)}{r} \right)^{-1}dr^2 +r^2 d\Omega^2,
\end{equation}
where $ d\Omega^2= d\theta^2 + \text{sin}^2 \theta d\phi^2 $, $ \Phi(r) $ and $ b(r) $ are the redshift function and shape function respectively. The minimal requirement of traversability of this particular wormhole geometry demands the following conditions:
\begin{enumerate}
	\item The wormhole is constructed by connecting two asymptotic flat regions at the throat. The throat radius is defined by a global minimum $ r=r_0 $, so the radial coordinate runs in the interval $ r \in [\ r_0,\infty )\ $.
	\item The redshift function $ \Phi(r) $ must be finite everywhere in order to avoid the presence of horizons and singularities. So, $ e^{\Phi(r)}>0 $ everywhere for $ r>r_0 $.
	
	In this context, the ultrastatic wormhole is a particular point of interest which defines the zero-tidal-force wormhole. Here, $ \Phi(r)=0 $, so that $ e^{\Phi(r)}=1 $ i.e., in a frame free from gravitational acceleration, a particle dropped from rest remains at rest \cite{morris1988,cataldo2017}.
	\item The flaring-out condition
	\begin{equation}
		\nonumber \frac{-rb'(r)+b(r)}{b^2(r)}>0,
	\end{equation}
	must hold at or near the throat $ r=r_0 $.
	\item The above mentioned conditions imply that $ b(r_0)=r_0 $ and $ b'(r_0)\le 1 $ for all $ r\ge r_0 $, where the equality of $ b'(r_0) $ only holds at the throat. Further, for $ r>r_0 \Rightarrow b(r)<r $. \label{con4}
	\item The asymptotic flatness implies that $ \Phi(r)\rightarrow 0 $ and $ b(r)/r \rightarrow 0 $ as $ r \rightarrow \infty $.
\end{enumerate}
 The redshift and shape function must obey these conditions (readers are referred to \cite{morris1988,tangphati2020,godani2020,samanta2020,tello2021}, for details).

Now, the Einstein tensor components for the spacetime metric are given by \cite{morris1988}
\begin{eqnarray}
	G_{tt} &=& \frac{b'}{r^2},
	\label{eq24}\\
	G_{rr} &=& -\frac{b}{r^3}+ 2\left( 1- \frac{b}{r} \right) \frac{\Phi'}{r},
	\label{eq25}\\
\nonumber	G_{\theta \theta} &=& G_{\phi \phi} =	\left( 1- \frac{b}{r} \right) \Big[ \Phi''+\Phi'^2 \\ &+& \left( \frac{-rb'+2r-b}{2r(r-b)} \right) \Phi' -\frac{rb'-b}{2r^2(r-b)} \Big].
	\label{eq26}
\end{eqnarray}

Therefore, using these components, the calculation of field equation (from Eq. \eqref{eq18}) of $ f(R,T) $ massive gravity is rather straightforward to obtain,
\begin{widetext}
	\begin{eqnarray}
	\rho &=& \frac{f}{16\pi}+ \frac{f_R}{8\pi} \left[ \left( 1- \frac{b}{r} \right) \left( \Phi''+\Phi'^2 \right) -\frac{rb'+3b-4r}{2r^2}\Phi' \right] -\left( 1- \frac{b}{r} \right) \frac{f_R''}{8\pi} + \left( \frac{rb'+3b-4r}{2r^2} \right) \frac{f_R'}{8\pi} + \left( \frac{2\gamma - \Lambda r}{8\pi r} \right),
	\label{eq27}
	\\
	\nonumber p_r &=& -\frac{f}{16\pi}+ \left( 1- \frac{b}{r} \right) \frac{f_R'' f_T}{8\pi(8\pi +f_T)} -\frac{f_R'}{(8\pi+f_T)} \left[ \left( \frac{rb'+3b-4r}{2r^2} \right) \frac{f_T}{8\pi}- \left( 1- \frac{b}{r} \right) \left( \Phi'+\frac2r \right) \right] 
	\\ 
	&-& \frac{f_R}{(8\pi+f_T)} \left[ \left( \frac{-rb'-3b+4r}{2r^2} \right) \Phi' \frac{f_T}{8\pi} + \left( \frac{-rb'+b}{2r^2} \right) \left( \Phi'+\frac2r \right) \right] - \left( 1-\frac{b}{r} \right) \left( \Phi''+\Phi'^2 \right) \frac{f_R}{8\pi}- \left( \frac{2\gamma - \Lambda r}{8\pi r} \right),
	\label{eq28} 
	\\
	\nonumber p_t &=& -\frac{f}{16\pi}+ \left( 1- \frac{b}{r} \right) \frac{f_R''}{8\pi} -\frac{f_R'}{(8\pi+f_T)} \left[ \left( \frac{rb'+3b-4r}{2r^2} \right) \frac{f_T}{8\pi} +\left( \frac{rb'+b-2r}{2r^2} \right) -\left( 1- \frac{b}{r} \right) \Phi' \right]
	\\
	&-& \frac{f_R}{(8\pi+f_T)} \left[ \left( \frac{r-b}{r^2}- \left( \frac{rb'+3b-4r}{2r^2} \right) \frac{f_T}{8\pi} \right)\Phi' - \frac{rb'+b}{2r^3} \right] -\frac{f_R f_T}{8\pi(8\pi+f_T)} \left( 1-\frac{b}{r}\right) \left( \Phi''+\Phi'^2 \right) - \left( \frac{2\gamma - \Lambda r}{8\pi r} \right).
	\label{eq29}
	\end{eqnarray}
\end{widetext}

Now, to construct wormhole solution, one can consider restricted choices of $ \Phi(r), ~b(r) $ and $ f(R,T) $ among others. The other way considers specific pressures (isotropic/anisotropic) or equations of state for $ p_r,~p_t $. Although, fixing the shape function prior to the solution may loose the effect of massive gravity on the shape. For an example, we can consider the wormhole solution in $ f(R) $ massive gravity by Tangphati {\it et. al.} \cite{tangphati2020}. In this present study, we will consider the anisotropic pressure solution in $ f(R,T) $ massive gravity and exhibit the effect of massive gravitons on the wormhole shape function. 

\section{Anisotropic Wormhole Solution}\label{aniso}
For the anisotropic pressure fluid wormhole solution in dRGT-$ f(R,T) $ massive gravity, we consider a simple choice of $ f(R,T) $, e.g., $ f(R,T)\equiv (\alpha R+\beta T) $ \footnote{Note that, this $ \alpha $ and $ \beta $ is different from the massive gravity $ \alpha $ and $ \beta $ parameters, and from here on, we will use $ \alpha $ and $ \beta $ only for the $ f(R,T) $ coefficients except for the deflection angle.}, where $ \alpha=1 $ and $ \beta=2 \lambda $ brings forth the well known linear choice of $ f(R,T) $ cosmology, i.e., $ f(R,T)\equiv (R+ 2\lambda T) $ \cite{harko2011}. We have $ f_R=\alpha, f_T=\beta $, and $ f_R'=f_R''=0 $ for the specific choice.  On the other hand, ultrastatic wormhole is chosen for the particular solution which results $ \Phi'(r)=\Phi''(r)=0 $, and we can directly compute the reduced EM components to get,
\begin{eqnarray}
	\rho &=& \frac{\alpha b'}{r^2 (8 \pi+\beta)}+ \frac{2 \gamma-\Lambda r}{2r (4\pi+\beta)},
	\label{eq30} \\
	p_r &=& -\frac{\alpha b}{r^3 (8 \pi+\beta)}- \frac{2 \gamma-\Lambda r}{2r (4\pi+\beta)},
	\label{eq31} \\
	p_t &=& \frac{\alpha(-r b'+b)}{2 r^3 (8 \pi+\beta)}- \frac{2 \gamma-\Lambda r}{2r (4\pi+\beta)}.
	\label{eq32}
\end{eqnarray}

The anisotropic pressure fluid equation is given by $ p_t=\sigma p_r $, where $ \sigma=1 $ defines the isotropic pressure condition (i.e., $ p_t=p_r=p $). At the same time, $ \sigma $ must be less than zero for anisotropic pressure fluid to ensure the asymptotic flatness of the spacetime.  Hence, for our particular choices, we can directly obtain the shape function by using equations \eqref{eq31} and \eqref{eq32}.
\begin{equation}\label{eq33}
	b(r)= \frac{r^2 (8\pi+ \beta)}{2 \alpha (4\pi + \beta)} \left[ \frac{-4 \gamma (1-\sigma) + \Lambda r (1-2 \sigma)}{(1-2 \sigma)} \right] + C r^{1+2 \sigma},
\end{equation}
where $ C $ is the constant of integration. It is expected to note that the shape function effectively depends on the massive gravity parameters ($ \gamma $ and $ \Lambda $), and hence on the mass of the graviton.

It is worth noting here that for $ \sigma=-1 $ and $ \gamma = \Lambda = 0 $, the equation reduces to the so called Ellis wormhole \cite{ellis1973} shape function.

However, one can obtain the position of wormhole throat from the root of $ \left( 1- \frac{b(r)}{r} \right) = 0 $, which (for our area of interest) eventually gives us
\begin{equation}\label{eq34}
	1+ A \gamma r - B \Lambda r^2 + C r^{2\sigma} = 0,
\end{equation}
where
\begin{equation}\label{eq35}
	A= \frac{2(8\pi +\beta)}{\alpha (4\pi + \beta)} \left( \frac{1-\sigma}{1-2 \sigma} \right),
	~~~~
	B= \frac{(8\pi + \beta)}{2 \alpha (4\pi +\beta)}.
\end{equation}

For Ellis wormhole, Eq. \eqref{eq34} becomes
\begin{equation}\label{eq36}
	1+ \bar{A} \gamma r - B \Lambda r^2 + \frac{C}{r^2} = 0.
\end{equation}

It is clear that the equation has four roots, and one of the real root determines the throat radius. One can easily impose condition \ref{con4} (from the 5 minimum requirements for wormhole geometry, section \ref{field_eq}) i.e., $ b(r=r_0) = r_0 $ to determine $ C $, so that the throat radius $ r_0 $ can be specified. Hence, the final shape function is given by
\begin{widetext}
	\begin{eqnarray}\label{eq37}
		b(r)= r \left( \frac{r}{r_0} \right)^{2 \sigma} - \frac{r (8\pi+ \beta)}{2 \alpha (4\pi + \beta) (1 - 2 \sigma)} \left[ 4 \gamma \left( 1-\sigma \right) \left( r - r_0 \left( \frac{r}{r_0} \right)^{2 \sigma} \right) - \Lambda \left( 1-2 \sigma \right) \left( r^2 - r_0^2 \left( \frac{r}{r_0} \right)^{2 \sigma} \right) \right].
	\end{eqnarray}
\end{widetext}

The result is much expected where massive gravity free (i.e., $ \gamma = \Lambda =0 $) expression turns out the usual solution of anisotropic pressure fluid in $ f(R,T) $ gravity. However, due to the presence of $ r $ and $ r^2 $ respectively with $ \gamma $ and $ \Lambda $ in the second term of the shape function changes the asymptotic structure of the spacetime. The effect is very much similar to the black hole solution in dRGT massive gravity. We refer \cite{panpanich2019} for the discussion in black hole asymptotic structure.
\begin{figure}[!]
	\centerline{\includegraphics[scale=.5]{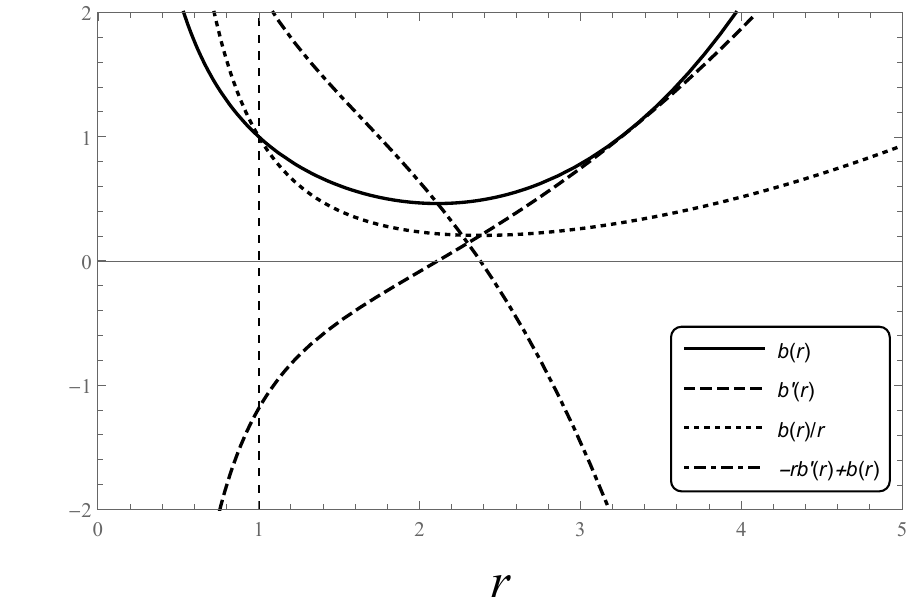}}
	\caption{Behaviour of the $ f(R,T) $-massive shape function against the Morris-Thorne wormhole properties for $ \alpha=1, \beta=2, \sigma=-1, \gamma=0.06 $, and $ \Lambda=0.07 $. The throat radius $ r_0=1 $ is exhibited by the vertical dashed line at $ r=1 $.}
	\label{shape_plot}
\end{figure}

To explore the shape function characteristics discussed in section \ref{field_eq}, we plot $ b(r), ~ b'(r), ~ b(r)/r $, and the flaring-out condition (i.e., $ -rb'(r) +b(r) $) in Fig. \ref{shape_plot}. It is observed that, though all the properties are satisfied near the throat, the conditions given by $ b'(r)<1, ~ b(r)/r <1 $ for all $ r>r_0 $, and $ b(r)/r \rightarrow 0 $ for $ r \rightarrow \infty $ are significantly violated after certain radial distance. So, it is straightforward to conclude that the asymptotic flatness of the spacetime vanishes after certain distance due to the presence of massive gravitons. We can interpret this as the result of repulsive effect of gravity, which is another unique feature of the dRGT massive gravity theory. A similar effect arises in the black hole solution in massive gravity where the asymptotic structure changes due to the said effect \cite{panpanich2019}. It should be noted here that repulsive behaviour of gravity is also found in the BHT massive gravity \cite{nakashi2019}, and in other modified gravities with exotic matters and energy \cite{kitamura2013,izumi2013,kitamura2014,nakajima2014,shaikh2017}.

Coming back to the discussion of non-asymptotic geometry, for $ r \rightarrow \infty $, $ b(r)/r $ did not approach $ '0' $, rather it goes to $ '\infty' $ due to the presence of linear $ r $ and $ r^2 $ terms with $ \gamma $ and $ \Lambda $ respectively in Eq. \eqref{eq37}, resulting the loss of asymptotic flatness. It is evident and can be exhibited mathematically that other choices of $ f(R,T) $ in ultrastatic wormhole geometry also involve such $ r $ and $ r^2 $ terms with $ \gamma $ and $ \Lambda $ in the shape function. Similar circumstances can be observed in other gravities such as Einstein's gravity in the background of massive gravitons (which will be discussed later in this literature), and in $ f(R) $ massive gravity wormholes \cite{tangphati2020}. Hence, we can conclude that it is a generic feature of wormholes in dRGT massive gravity.

Now, we need to check the presence of repulsive gravity effect in the wormhole solution to verify if it has any role on the asymptotic structure. To easily verify the phenomenon, one can introduce the photon deflection angle on the wormhole, which eventually goes negative in a spacetime if repulsive gravity acts on the photons \cite{panpanich2019}.

\subsection{Repulsive Behaviour of Gravity}\label{repul_grav}
To check the deflection angle of photons from null geodesics, we first introduce a general spherically symmetric and static line element given by \cite{misner1973, schutz2014}
\begin{equation}\label{eq38}
	ds^2 = -A(r) dt^2 +B(r) dr^2 +C(r) d\Omega^2.
\end{equation}

The geodesic equation, which relates the momenta one-forms of a freely falling body and background geometry is given by \cite{schutz2014}
\begin{equation}\label{eq39}
	\frac{dp_\beta}{d\lambda} = \frac{1}{2} g_{\nu \alpha, \beta} p^\nu p^\alpha ,
\end{equation}
where $ \lambda $ is the affine parameter. One can immediately tell that if the components of $ g_{\alpha \nu} $ are independent of $ x^\beta $ for a fixed index $ \beta $, then $ p_\beta $ is a constant of motion. Hence, if we consider only the equatorial slice by setting $ \theta = \pi/2 $, then all the $ g_{\alpha \beta} $ are independent of $ t, \theta, \phi $ in Eq. \eqref{eq39}, i.e., one can obtain the respective killing vector fields $ \delta^{\mu}_{\alpha} \partial_\nu $ with $ \alpha $ as a cyclic coordinate. Now, we can set the constants of motion $ p_t $ and $ p_\phi $ as
\begin{equation}\label{eq40}
	p_t = -E, ~~~~~~~~~~~~~ p_\phi = L,
\end{equation}
where $ E $ and $ L $ are the energy and angular momentum of the photon respectively. Thus, we have
\begin{eqnarray}\label{eq41}
\nonumber	p_t = \dot{t} = g^{t \nu} p_\nu = \frac{E}{A(r)}, \\
	p_\phi = \dot{\phi} = g^{\phi \nu} p_\nu = \frac{L}{C(r)},
\end{eqnarray}
where the overdot represents the differentiation w.r.t. affine parameter $ \lambda $. Again, the radial null geodesic can be obtained easily as
\begin{equation}\label{eq42}
	\dot{r}^2 = \frac{1}{B(r)} \left( \frac{E^2}{A(r)} - \frac{L^2}{C(r)} \right).
\end{equation}

However, one can write the equation for the photon trajectory in terms of impact parameter $ \mu = L/E $, as
\begin{equation}\label{eq43}
	\left( \frac{dr}{d\phi} \right)^2 = \frac{C(r)^2}{\mu^2 B(r)} \left[ \frac{1}{A(r)} - \frac{\mu^2}{C(r)} \right].
\end{equation}

Now, one can obtain the deflection angle of photon by considering a source of photon radius $ r_s $ causing the geometry, then the photons can hit the surface only when an existing solution $ r_0 $ obeys the condition $ r_0 >r_s $, and $ \dot{r}^2 =0 $. Here, $ r_0 $ is the distance of closest approach or turning point. In that case, the impact parameter becomes
\begin{equation}\label{eq44}
	\mu = \frac{L}{E} = \pm \sqrt{\frac{C(r_0)}{A(r_0)}},
\end{equation}
and it is obvious that for weak gravity limit, $ \mu \approx \sqrt{C(r_0)} $. Thus, if a photon is coming from polar coordinate limit given by $ \lim\limits_{r \rightarrow \infty} \left( r, -\frac{\pi}{2}-\frac{\alpha}{2} \right) $, passes through the turning point at $ (r_0, 0) $ and approaches $ \lim\limits_{r \rightarrow \infty} \left( r, \frac{\pi}{2}+\frac{\alpha}{2} \right) $, then the deflection angle of photon is defined as this $ \alpha $, which is a function of $ r_0 $ \cite{bhattacharya2010}. We can compute it from Eq. \eqref{eq43} as
\begin{equation}\label{eq45}
	\alpha(r_0) = -\pi + 2 \int_{r_0}^{\infty}
	\frac{\sqrt{B(r)} dr}{\sqrt{C(r)} \left[ \left( \frac{A(r_0)}{A(r)} \right) \left( \frac{C(r)}{C(r_0)} \right) -1 \right]^{1/2}}.
\end{equation}

For the choices of metric coefficients in wormhole geometry, the deflection angle becomes
\begin{equation}\label{eq46}
	\alpha(r_0)=-\pi+2 \int_{r_0}^{\infty} \frac{dr}{r \left[ \left( 1- \frac{b(r)}{r} \right) \left( \frac{r^2}{r_0^2} -1 \right) \right]^{1/2} }.
\end{equation}

\begin{figure}[!]
	\centerline{\includegraphics[scale=.6]{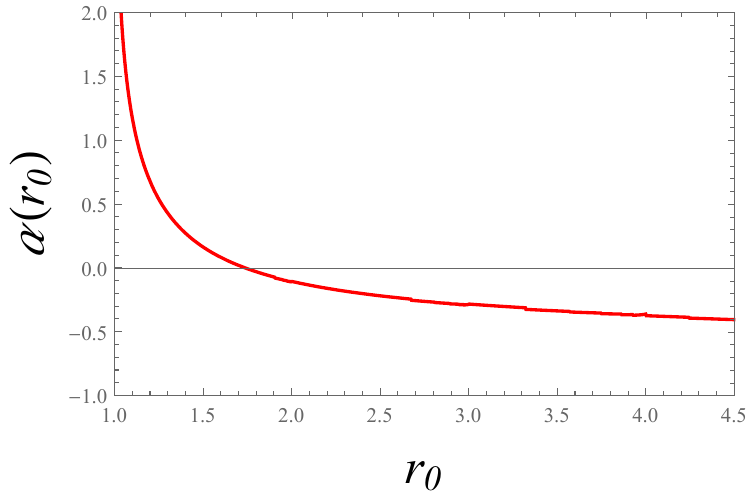}}
	\caption{Numerical result of the deflection angle of photons in $ f(R,T) $-massive gravity for $ r_0=1, \sigma=-1, \gamma=0.06, \Lambda=0.07, \alpha=1 $, and $ \beta=2 $. The plot exhibit the negative deflection angle representing the repulsive nature of gravity. It also verifies the reason of non-asymptotic behaviour.}
	\label{deflec_plot}
\end{figure}

One can now readily exhibit the deflection angle of photons in massive gravity by numerically integrating the above equation after imposing the shape function given by Eq. \eqref{eq37}, which is shown in Fig \ref{deflec_plot}.

It is interesting to observe that the deflection angle becomes negative after a certain value of $ r_0 $. We can interpret this as the effect of repulsive gravity, where neglecting massive gravity parameters $ \gamma $ and $ \Lambda $ (i.e. neglecting massive gravity), one can comfortably verify the non-existence of negative deflection angle \cite{mishra2018}. At the same time, referring Fig. \ref{shape_plot}, it is verified that the inconsistencies in the spacetime structure like non-asymptotic flatness occurs beyond the radial distance where the repulsive effect of gravity dominates. So, we can conclude that the loosening of asymptotic structure results from the repulsive behaviour of gravity.

However, the repulsive anisotropic dark energy nature is inherent in massive gravitons. So, their presence may physically cause the repulsive gravity effect in dRGT massive gravity.

\section{Wormhole in Einstein-massive gravity}\label{einstein}
To start the discussion of Einstein gravity with the background of dRGT massive gravity, the action is considered as
\begin{equation}\label{ein1}
	S = \int d^4 x \sqrt{-g} \bigg(\frac{1}{16\pi}\Big[ R + m^2_g \mathcal{U}(g,\phi^a) \Big] +\mathcal{L}_{m} \bigg).
\end{equation}

We can now vary the action with respect to $ g_{\mu \nu} $ after considering the massive gravity and other parameters from section \ref{field_eq} and get
\begin{equation}\label{ein2}
	G_{\mu \nu}=  8\pi T_{\mu \nu}- m_g^2 X_{\mu \nu}.
\end{equation}

However, one can easily derive the above results just by considering $ f(R,T)=R $, i.e., $ f_R(R,T)=1 $, in $ f(R,T) $ massive gravity.

In Eq. \eqref{ein2}, the EM tensor of the Einstein gravity follows the energy conservation law, and the total EM tensor defined by the coupled perfect fluid-massive graviton system is given by
\begin{equation}\label{ein3}
	T^{tot}_{\mu \nu}= diag\left( -\rho^{E}-\rho^{(g)}, p_r^{E}+p_r^{(g)}, p_t^{E}+p_t^{(g)}, p_t^{E}+p_t^{(g)} \right),
\end{equation}
where $ `E' $ in the superscript represents the EM components for Einstein gravity. So, one can directly write $ \nabla^\mu T^{tot}_{\mu \nu}=0 $, $ \nabla^\mu T_{\mu \nu}=0 $ and $ \nabla^\mu T^{(g)}_{\mu \nu}=0 $. In Einstein-massive gravity, the construction of wormhole can be possible with non-exotic matter source, where the violation of energy conditions may manipulated by the exotic nature of massive gravitons.

Now, to derive the calculations of field equation for Morris-Thorne wormhole, we may use Eq. \eqref{eq24}, \eqref{eq25}, and \eqref{eq26} in Eq. \eqref{ein2} and obtain,
\begin{eqnarray}
	\rho &=& \frac{b'}{8 \pi r^2}+\left( \frac{2 \gamma-\Lambda r}{8 \pi r} \right),
	\label{ein4} \\
	p_r &=& -\frac{b}{8 \pi r^3}+\left( 1-\frac{b}{r} \right) \frac{\Phi'}{4\pi r}-\left( \frac{2 \gamma-\Lambda r}{8 \pi r} \right),
	\label{ein5} \\
	p_t &=& \left(\frac{-r b'+b}{16\pi r^3}\right) + \left(\frac{-rb'+2r-b}{16\pi r^2}\right) \Phi' \nonumber \\
	&+& \frac{1}{8\pi} \left( 1-\frac{b}{r} \right) \left(\Phi'' + \Phi'^2 \right) -\left( \frac{2 \gamma-\Lambda r}{8 \pi r} \right).
	\label{ein6]}
\end{eqnarray}

Here, we are about to construct an anisotropic wormhole solution with restricted choice of redshift function, e.g. $ \Phi= $ constant (i.e. $ \Phi'=\Phi''=0 $), which represents the ultrastatic wormhole geometry. Hence, the equation of state for the anisotropic pressure solution, $ p_t=\sigma p_r $ provides the shape function of the wormhole. By using the boundary condition at the throat junction, i.e. $ b(r_0)=r_0 $, the final shape function is given by
\begin{widetext}
	\begin{eqnarray}\label{ein7}
		b(r)= r \left[ \left( \frac{r}{r_0} \right)^{2 \sigma} - \frac{4 \gamma (1-\sigma)}{(1-2 \sigma)} \left( r - r_0 \left( \frac{r}{r_0} \right)^{2 \sigma} \right) + \Lambda \left( r^2 - r_0^2 \left( \frac{r}{r_0} \right)^{2 \sigma} \right) \right].
	\end{eqnarray}
\end{widetext}

\begin{figure}[!]
	\centerline{\includegraphics[scale=.5]{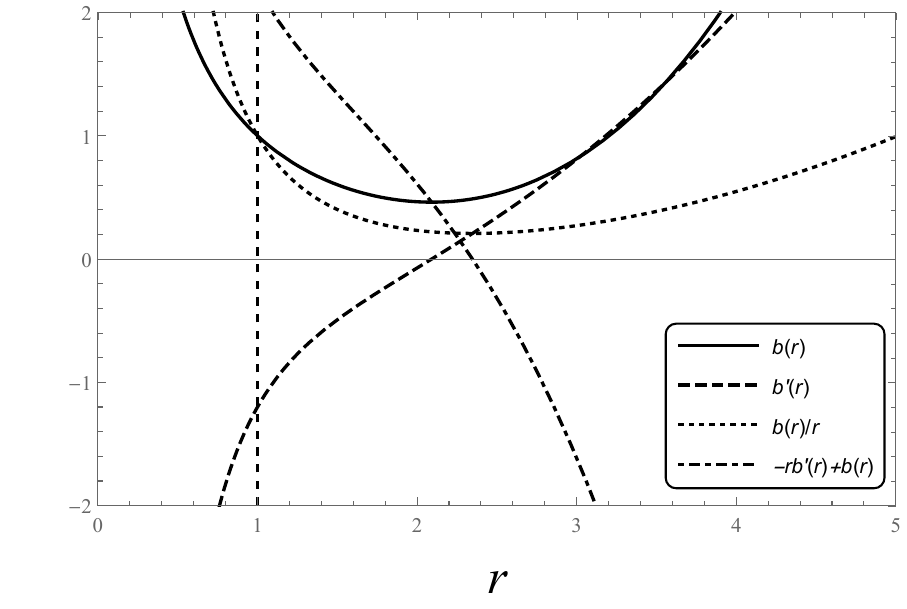}}
	\caption{Behaviour of the Einstein-massive shape function against the Morris-Thorne wormhole properties for $ \sigma=-1, \gamma=0.06 $, and $ \Lambda=0.07 $. The throat radius $ r_0=1 $ is exhibited by the vertical dashed line at $ r=1 $.}
	\label{shape_ein}
\end{figure}

The shape function is very much similar to that obtained in $ f(R,T) $ gravity, and the visualization of the Morris-Thorne type shape function properties, i.e. $ b(r),~b'(r),~b(r)/r, $ and $ (-rb'(r)+b(r)) $ are plotted in Fig. \ref{shape_ein}. The plots violate the asymptotic flatness after a certain radial distance, as it also contains linear $ r $ and $ r^2 $ terms with $ \gamma $ and $ \Lambda $ respectively. However, if the expression is freed from the massive gravity parameters, i.e. $ \gamma=\Lambda=0 $, asymptotically well-behaved shape function is recovered.

\begin{figure}[!]
	\centerline{\includegraphics[scale=.6]{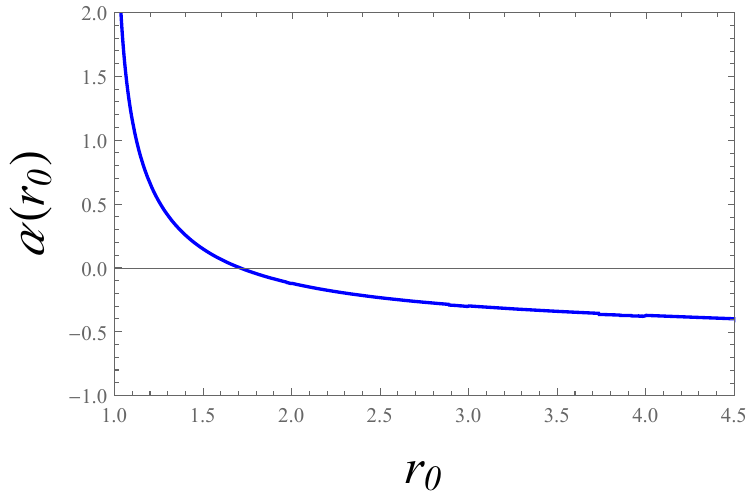}}
	\caption{Numerical result of the deflection angle of photons in Einstein-massive gravity for $ r_0=1, \sigma=-1, \gamma=0.06, \Lambda=0.07 $. The plot exhibit the negative deflection angle representing the repulsive nature of gravity. It also verifies the reason of non-asymptotic behaviour.}
	\label{deflec_ein}
\end{figure}

The discussion of repulsive effect of gravity in this ground is indeed necessary, and we directly use Eq. \eqref{eq46} to plot the numerical solution of deflection angle of photons in dRGT extension of Einstein gravity. The existence of negative deflection angle indicate the presence of repulsive gravity, where all the other comprehensive discussions are similar to the $ f(R,T) $-massive gravity. The plot is exhibited in Fig. \ref{deflec_ein}, and comparing it with Fig. \ref{shape_ein}, it is satisfied that the Morris-Thorne wormhole shape function properties are violated after the radial distance where the deflection angle goes negative. So, also in Einstein-massive gravity, the asymptotic flatness of the traversable wormhole is lost due to the repulsive effect of gravity, and it is evident that it arises from the anisotropic dark energy nature (repulsive nature) of massive gravitons.

\section{Energy Conditions}\label{en_con}
Discussing energy conditions in traversable wormhole configuration is highly important to have an idea of the matter content on the wormhole. As discussed, the throat of usual wormholes must be threaded by a matter of negative energy density that violates the Null energy condition (NEC) and is termed as `exotic matter'. The matter is used to keep the wormhole throat open, thus making it traversable.

Here, in this section, we will discuss the energy conditions for wormhole configuration consecutively in $ f(R,T) $-massive gravity and Einstein-massive gravity.

From the definitions and mathematical fundamentals, one can summarize the energy conditions in principal pressure forms which is given by
\\
(i) NEC : $ \rho+p_r\ge0, ~~ \rho+p_t\ge0 $;\\
(ii) WEC : $ \rho\ge0, ~~ \rho+p_r\ge0, ~~ \rho+p_t\ge0 $;\\
(iii) SEC : $ \rho+p_r\ge 0, ~~ \rho+p_t\ge0, ~~ \rho+p_r+2p_t \ge 0 $;\\
(iv) DEC : $ \rho\ge0, ~~ \rho-|p_r|\ge0, ~~ \rho-|p_t|\ge0 $.

\begin{figure*}[t]
	\centering
	\subfloat[]{{\includegraphics[width=8cm]{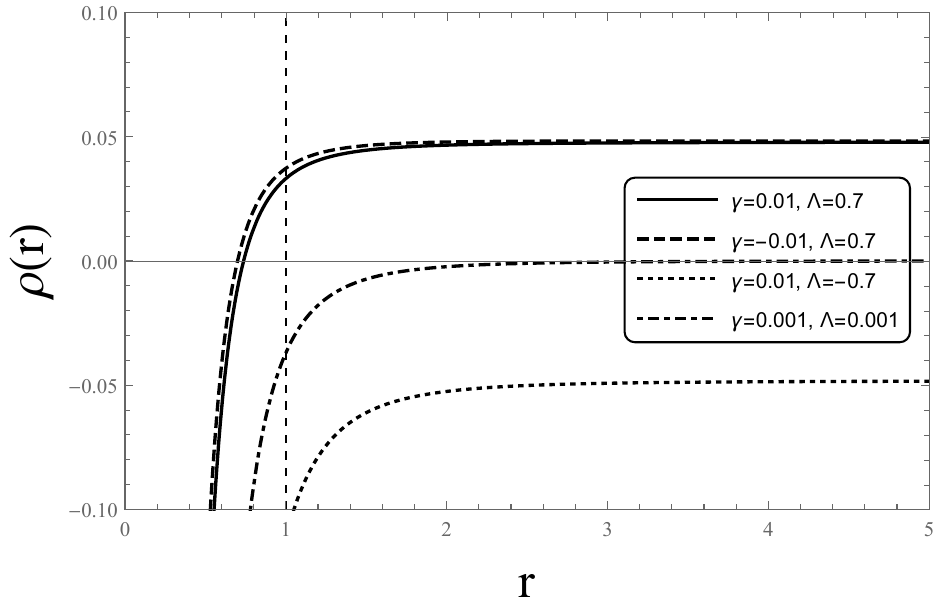}}}\qquad
	\subfloat[]{{\includegraphics[width=8cm]{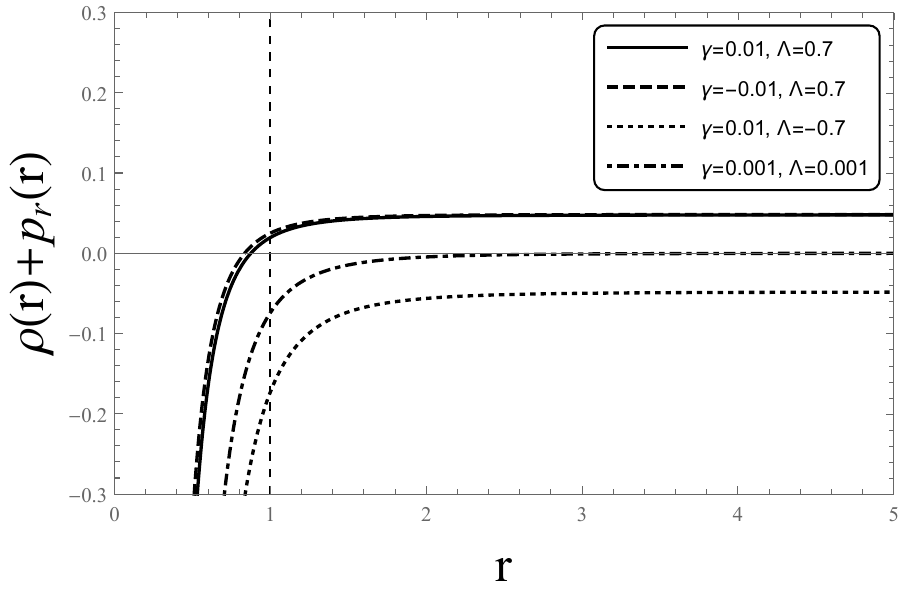}}}\\
	\subfloat[]{{\includegraphics[width=8cm]{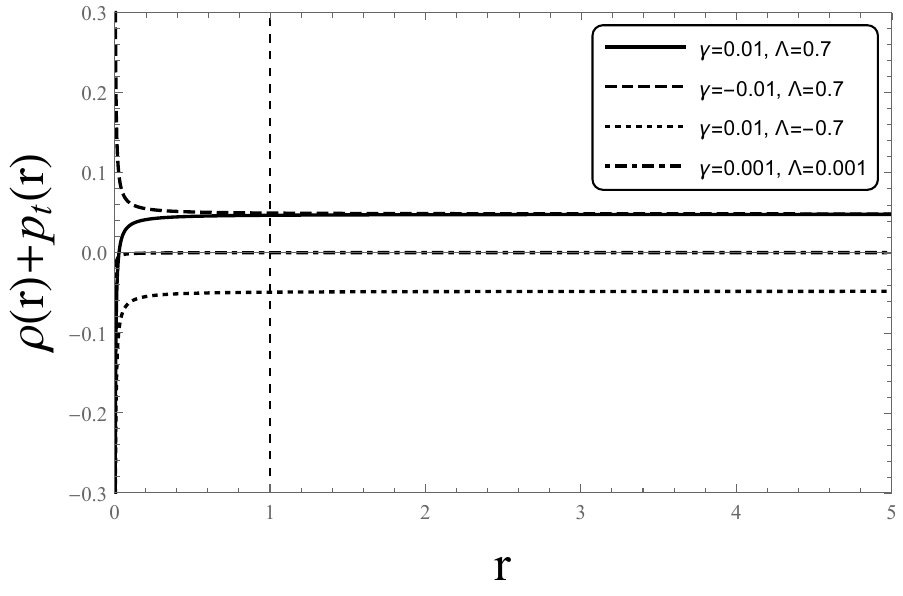}}}\qquad
	\subfloat[]{{\includegraphics[width=8cm]{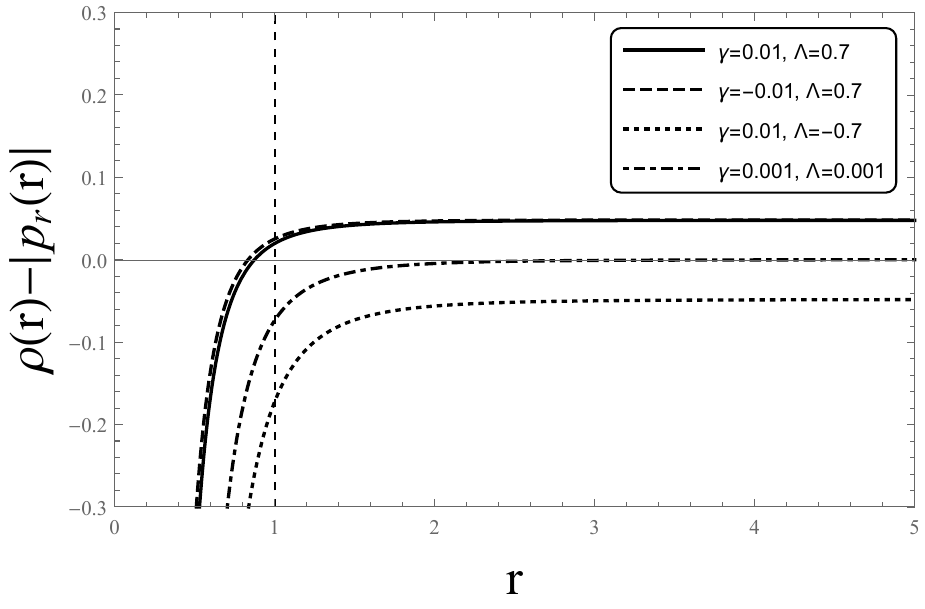}}}\\	\subfloat[]{{\includegraphics[width=8cm]{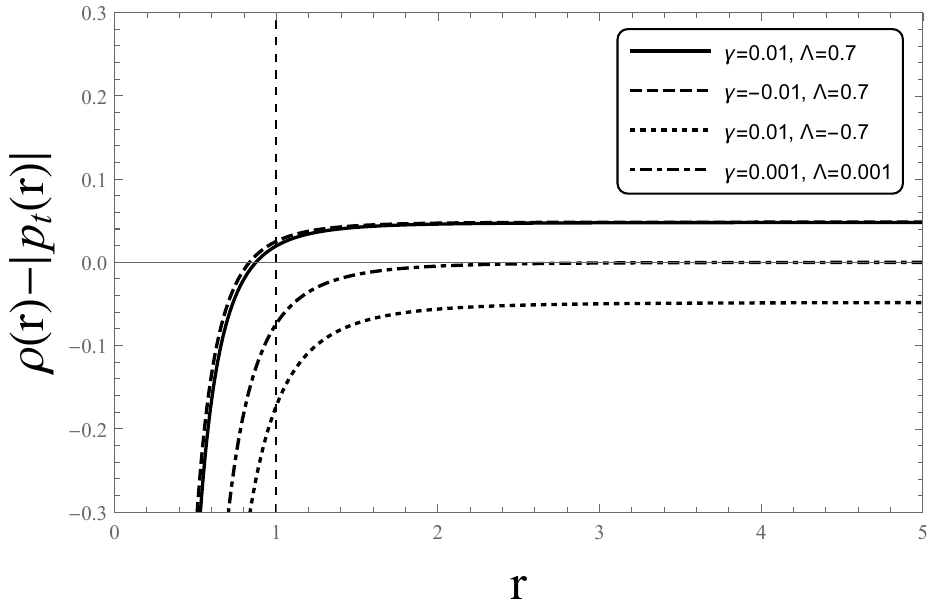}}}\qquad
	\subfloat[]{{\includegraphics[width=8cm]{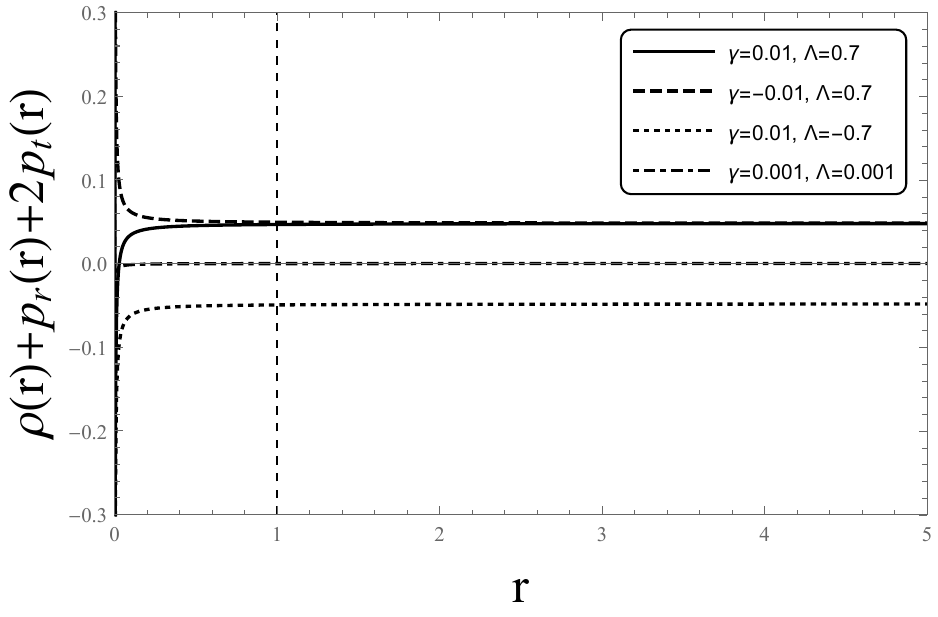}}}
	\caption{Plots demonstrating the variation of $ \rho, ~ \rho+p_r, ~ \rho+p_t, ~ \rho-|p_r|, ~ \rho-|p_t| $, and $ \rho+p_r+2p_t $ with radial distance $ r $. The nature of the plots are obtained for four different combinations of $ \gamma $ and $ \Lambda $, considering $ \alpha=1, ~ \beta=2, ~ \sigma=-1, $ and $ r_0=1 $. The vertical dashed line shows the position of the wormhole throat.}
	\label{en_con_plot}
\end{figure*}

Now, one can use standard mathematical tools to calculate $ \rho, ~ \rho+p_r, ~ \rho+p_t, ~ \rho-|p_r|, ~ \rho-|p_t|, $ and $ \rho+p_r+2p_t $ from Eq. \eqref{eq30}, \eqref{eq31}, and \eqref{eq32} by imposing the shape function of Eq. \eqref{eq37}. Using these six terms, all the four energy conditions for $ f(R,T) $-massive gravity can be investigated. The mathematical results of the terms are plotted in Fig. \ref{en_con_plot} for $ \alpha=1, ~ \beta=2, ~ \sigma=-1, $ and $ r_0=1 $. Alongside, the summaries of the results from the plots are listed in Table \ref{en_con_table}.
\begin{table*}[t]
	\centering
	\begin{tabular}{ p {2.5 cm} p {4 cm} p {4.5 cm} }
	\hline
	\hline
	~~Terms			&	$ \gamma $ and $ \Lambda $			&	Result \\
	\hline
	~~$ \rho $		&	$ \gamma=0.01, ~ \Lambda=0.7 $		&	$ \ge 0 $ for $ r \in [\ 0.73,\infty )\ $\\
					&	$ \gamma=-0.01, ~ \Lambda=0.7 $		&	$ \ge 0 $ for $ r \in [\ 0.69,\infty )\ $\\
					&	$ \gamma=0.01, ~ \Lambda=-0.7 $		&	Always $ <0 $\\
					&	$ \gamma=0.001, ~ \Lambda=0.001 $	&	$ \ge 0 $ for $ r \in [\ 5.29,\infty )\ $\\
	\hline
	~~$ \rho+p_r $	&	$ \gamma=0.01, ~ \Lambda=0.7 $		&	$ \ge 0 $ for $ r \in [\ 0.87,\infty )\ $\\
					&	$ \gamma=-0.01, ~ \Lambda=0.7 $		&	$ \ge 0 $ for $ r \in [\ 0.83,\infty )\ $\\
					&	$ \gamma=0.01, ~ \Lambda=-0.7 $		&	Always $ <0 $\\
					&	$ \gamma=0.001, ~ \Lambda=0.001 $	&	$ \ge 0 $ for $ r \in [\ 6.09,\infty )\ $\\	
	\hline
	~~$ \rho+p_t $	&	$ \gamma=0.01, ~ \Lambda=0.7 $		&	$ \ge 0 $ for $ r \in [\ 0.02,\infty )\ $\\
					&	$ \gamma=-0.01, ~ \Lambda=0.7 $		&	Always $ >0 $ \\
					&	$ \gamma=0.01, ~ \Lambda=-0.7 $		&	Always $ <0 $\\
					&	$ \gamma=0.001, ~ \Lambda=0.001 $	&	$ \ge 0 $ for $ r \in [\ 2,\infty )\ $\\
	\hline
	~~$\rho-|p_r| $	&	$ \gamma=0.01, ~ \Lambda=0.7 $		&	$ \ge 0 $ for $ r \in [\ 0.87,\infty )\ $\\
					&	$ \gamma=-0.01, ~ \Lambda=0.7 $		&	$ \ge 0 $ for $ r \in [\ 0.83,\infty )\ $\\
					&	$ \gamma=0.01, ~ \Lambda=-0.7 $		&	Always $ <0 $\\
					&	$ \gamma=0.001, ~ \Lambda=0.001 $	&	$ \ge 0 $ for $ r \in [\ 6.09,\infty )\ $\\	
	\hline
	~~$\rho-|p_t| $	&	$ \gamma=0.01, ~ \Lambda=0.7 $		&	$ \ge 0 $ for $ r \in [\ 0.87,\infty )\ $\\
					&	$ \gamma=-0.01, ~ \Lambda=0.7 $		&	$ \ge 0 $ for $ r \in [\ 0.83,\infty )\ $\\
					&	$ \gamma=0.01, ~ \Lambda=-0.7 $		&	Always $ <0 $\\
					&	$ \gamma=0.001, ~ \Lambda=0.001 $	&	$ \ge 0 $ for $ r \in [\ 6.09,\infty )\ $\\	
	\hline
~~$ \rho+p_r+2p_t $	&	$ \gamma=0.01, ~ \Lambda=0.7 $		&	$ \ge 0 $ for $ r \in [\ 0.02,\infty )\ $\\
					&	$ \gamma=-0.01, ~ \Lambda=0.7 $		&	Always $ >0 $ \\
					&	$ \gamma=0.01, ~ \Lambda=-0.7 $		&	Always $ <0 $\\
					&	$ \gamma=0.001, ~ \Lambda=0.001 $	&	$ \ge 0 $ for $ r \in [\ 2,\infty )\ $\\
	\hline
	\hline
	\end{tabular}
	\caption{Numerical results for the possible regions where respective energy conditions are satisfied.}
	\label{en_con_table}
\end{table*}

From Fig. \ref{en_con_plot} and Table \ref{en_con_table}, we can conclude that all the energy conditions including the energy density are satisfied throughout the wormhole from $ r=0.87 $ and $ r=0.83 $, respectively for $ \gamma=0.01, ~ \Lambda=0.7 $ and $ \gamma=-0.01, ~ \Lambda=0.7 $. On the other hand, one can also conclude that the energy condition components does not differ much for the signs of $ \gamma $. However, they are highly dependent on the effective cosmological constant ($ \Lambda $). For $ \gamma=0.01, ~ \Lambda=-0.7 $, all the components are completely negative, resulting violation of energy conditions. Alongside, for smaller $ \gamma $ and $ \Lambda $ terms, i.e., for less effects of massive gravity, the energy conditions are violated at the wormhole throat, but gets satisfied after a certain radial distance.

Therefore, it is interesting to note that, for the first two choices of $ \gamma $ and $ \Lambda $, i.e., $ \gamma=0.01, ~ \Lambda=0.7 $ and $ \gamma=-0.01, ~ \Lambda=0.7 $, the wormhole can be constructed with ordinary matter, which satisfies all the energy conditions. The situation is investigated in \cite{epl_paper} with great details, where it is established that we can have wormholes with ordinary matter in modified theories of gravity. However, the violation of NEC is customary at the wormhole throat to keep it open for traversability. They investigated that the presence of effective geometric matter (e.g. the curvature EM component in $ f(R,T) $ gravity) can play the role of energy condition violation. Few of the recent studies in $ f(R,T) $ gravity \cite{banerjee2021,ilyas2022} deal with the wormhole models with non-exotic matters, where the curvature matter source acts as the exotic matter and performs in the violation of energy conditions. Here, in this current scenario, the matter source of massive gravitons (which acts as the anisotropic dark energy) added with the curvature matter source, violate the energy condition components. So, the usual matter source can be ordinary matter. However, the repulsive dark energy nature in massive gravitons is so strong that it brings forth the repulsive gravity effect in the spacetime and pushes the geometry so strongly that the asymptotic flatness is effected.

On the other hand, the Einstein-massive gravity does not differ much from our linear $ f(R,T) $-massive gravity model. By imposing $ \alpha=1, $ and $ \beta=0 $ in Eq. \eqref{eq30}, \eqref{eq31}, \eqref{eq32}, and \eqref{eq37}, we can readily recover EM components and the shape function for Einstein gravity. Mathematically, the EM components are very loosely dependent on the $ \beta $ values, and graphically, the energy conditions possess adequately similar behaviour for $ \alpha=1, $ and $ \beta=0 $. For $ \gamma=0.01, ~ \Lambda=0.7 $ and $ \gamma=-0.01, ~ \Lambda=0.7 $, it also contains ordinary matter wormhole. But here, only the massive gravity sector adds the energy condition violations. So, it is evident that the discussions for Einstein-massive gravity model are similar to that of the $ f(R,T) $ gravity.

\section{Equilibrium Analysis}\label{stability}
The equilibrium condition for our present work is provided by the generalized Tolman–Oppenheimer–Volkov (TOV) equation which is given by
\begin{equation}
	\frac{dp_r}{dr}+\frac{\Phi'}{2}(\rho+p_r)+\frac2r(p_r-p_t)=0.
	\label{tov1}
\end{equation}

The above equation is an important and elegant method and can be used to examine the stability condition of the astrophysical solutions including wormhole/compact objects. Readers may refer to \cite{banerjee2021,oppenheimer1939,tolman1939,leon1993,jawad2015,sokoliuk2021} for detailed study.

The Eq. \eqref{tov1} can also be written as
\begin{equation}
	F_a+F_g+F_h=0,
	\label{tov2}
\end{equation}
which provides the equilibrium condition for the wormhole. Here,
\begin{eqnarray}
	F_a &=& \frac2r(p_t-p_r), \nonumber \\
	F_g &=& -\frac{\Phi'}{2}(\rho+p_r), \nonumber \\
	F_h &=& -\frac{dp_r}{dr}. \nonumber
\end{eqnarray}
where, $ F_a $ denotes the force due to the anisotropic matter of the wormhole, $ F_g $ is the gravitational force, and $ F_h $ is the hydrostatic force. $ F_a $ arises due to the modification of the gravitational Lagrangian of the Einstein-Hilbert action. However, it is clear from Eq. \eqref{tov2} that for the system to be in equilibrium, the sum of the three different forces must be equal to zero.
\begin{figure}[!]
	\centerline{\includegraphics[scale=.65]{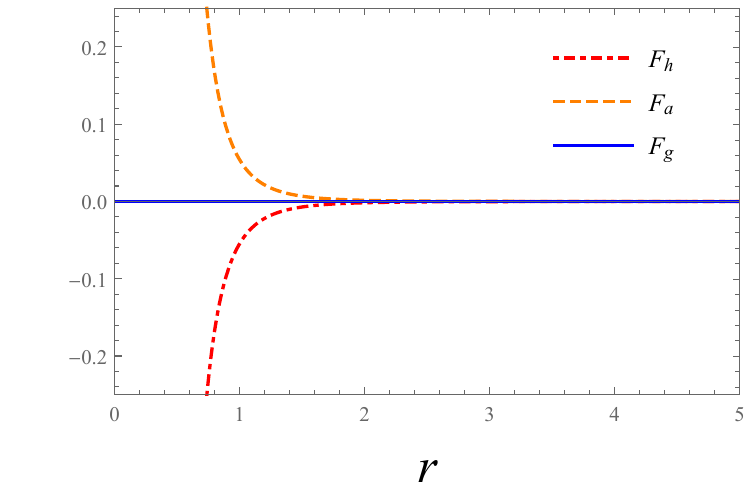}}
	\caption{The three forces for the equilibrium condition are plotted against $ r $, for $ r_0=1, \sigma=-1, \gamma=0.06, \Lambda=0.07 $ and $ \alpha=1, \beta=2 $. Here $ F_g=0 $ for the ultrastatic wormhole choice.}
	\label{fig_tov}
\end{figure}

In Fig. \ref{fig_tov}, the behaviour of $ F_a,~F_g, $ and $ F_h $ are shown for a particular choice of the parameter values where the energy conditions are satisfied. However, the value of $ F_g $ is zero due to the assumption of $ \Phi $ to be constant (i.e. the gravitational force has no effect on our model). From the figure, one can visualize that the other two forces are exactly same and opposite to each other. Thus, it is evident that the equilibrium of forces is achieved due to the combined effect of the three force terms, and hence this supports the stability of the system.

\section{Discussions}\label{discuss}
In this work, we have presented the wormhole solution in the dRGT massive gravity extension of $ f(R,T) $ gravity and Einstein gravity. For the anisotropic pressure wormhole solution, we considered a linear choice of $ f(R,T) $, i.e., $ f(R,T)\equiv \alpha R + \beta T $, which is a straightforward choice of the linear $ f(R,T) $ cosmology model, i.e., $ f(R,T)\equiv R+ 2 \lambda T $. One of the advantages in this particular choice is that, one can easily reduce this to Einstein gravity by considering $ \alpha=1 $ and $ \beta=0 $. On the other hand, from this model, $ f(R) $-massive gravity can also be recovered by suitable choices of parameters. Although, in $ f(R,T) $ gravity, the energy conservation law is violated \cite{harko2011}, but still wormhole solutions (also with non-exotic matter) can be constructed in this theory. For example, readers can go through the following literature \cite{banerjee2021,ilyas2022}. However, one may also consider conservation of energy density, where interestingly the field equations remains similar \cite{chakraborty2013}, and the possibility of non-exotic matter wormhole can be investigated.

The shape function obtained from the anisotropic pressure solution in the model involves product terms of $ r $ and $ r^2 $ respectively with the massive gravity parameters $ \gamma $ and $ \Lambda $, which causes non-asymptotic flat geometry in the spacetime after a certain radial distance. Although, we can extend the region of asymptotically well-behaved spacetime by considering smaller values of $ \gamma $ and $ \Lambda $. However, a globally complete asymptotic flat wormhole solution can be achieved by imposing restricted choices of shape function. The non-asymptotic flatness and the reason is similar to the black hole solution in massive gravity where $ \gamma r $ and $ \Lambda r^2 $ terms arise in the metric coefficients \cite{panpanich2019}.

The repulsive effect of gravity in a spacetime can be examined by the negative deflection angle of photons. It is investigated that, in an extended spacetime description with massive gravity, the deflection angle goes negative due to the terms consisting of $ \gamma $ and $ \Lambda $. Hence, the non-asymptotic flatness and the presence of repulsive gravity is the generic feature of dRGT massive gravity, and the massive gravitons source the effects both in the black holes and wormholes.

On the other hand, if we investigate the equation of state ($ p_r = \omega \rho $) to obtain the shape function, we get
\begin{widetext}
	\begin{eqnarray}
\nonumber		b(r)= r \left( \frac{r_0}{r} \right)^{1+ 1/ \omega} + \frac{r (8\pi+ \beta) (1+ \omega)}{2 \alpha (4\pi + \beta)} \left[ - \frac{2 \gamma}{1 + 2 \omega} \left( r - r_0 \left( \frac{r_0}{r} \right)^{1+ 1/ \omega} \right) + \frac{\Lambda}{1 + 3 \omega} \left( r^2 - r_0^2 \left( \frac{r_0}{r} \right)^{1+ 1/ \omega} \right) \right].
	\end{eqnarray}
\end{widetext}
One can readily verify that the expression is very similar to the anisotropic pressure solution, so it also replicates all the properties of the anisotropic solution. 

The dRGT-Einstein gravity is the generalization of $ f(R,T) $ model with $ \alpha=1 $, and $ \beta=0 $, so it reproduces all the effects and analysis present in the $ f(R,T) $ model. The shape function violates the Morris-Thorne properties, and the asymptotic flatness is lost just beyond the radial distance where the repulsive gravity effect arises due to the massive gravitons.

The discussion of wormhole solution in massive gravity has a particular point of interest as the energy-momentum tensor arising from the massive gravitons naturally fulfil the violation of null energy conditions. Further, the introduction of $ f(R,T) $ modification in massive gravity is more viable to construct wormholes with well-behaved matter source. In wormholes of modified gravity, there are two kinds of matter sources, one of which is the usual matter and another is the effective geometric matter which arises from the geometry of modified gravity \cite{epl_paper}. In the current study, it is established that for different values of $ \gamma $ and $ \Lambda $, there are large number of possibilities of ordinary matter wormhole that satisfies all the energy conditions. Here, the usual matter source of wormhole is ordinary matter, where the matter source of massive gravitons coupled with the curvature term of $ f(R,T) $ gravity act as the geometric matter that sources the negative energy density necessary to keep the wormhole throat open. It is due to the exotic anisotropic dark energy nature of coupled curvature and massive gravitons. But for Einstein-massive gravity, it is only the massive gravitons that sources the exotic component in non-exotic matter wormhole. In this context, it is customary to note that the strong repulsive nature of massive gravitons give rise to the repulsive effect of gravity and pushes the geometry for non-asymptotic flatness, both in $ f(R,T) $ and Einstein massive gravity.

Finally, the Tolman–Oppenheimer–Volkov (TOV) equation is investigated for the equilibrium condition analysis. Although, the gravitational force is zero due to the constant redshift function, but the wormhole model is stable by the interaction of anisotropic and hydrostatic force terms.

\section*{Acknowledgement}
The authors thank the anonymous referee whose comments and suggestions improved the quality and visibility of the paper. S.C. thanks FIST program of DST, Department of Mathematics, JU (SR/FST/MS-II/2021/101(C)).

\end{document}